\documentclass[namedreferences]{solarphysics}

\usepackage[hyperref,optionalrh,showbiblabels]{spr-sola-addons} 

\usepackage{amsmath,amssymb}    
\usepackage{cases}
\usepackage{graphicx}        
\usepackage{color}           
\usepackage{breakurl}        

\newcommand{\etal}{{\it et al.}}

\newcommand{\adv}{    {\it Adv. Space Res.}}

\newcommand{\aap}{    {\it Astron. Astrophys.}}

\newcommand{\apj}{    {\it Astrophys. J.}}
\newcommand{\apjl}{   {\it Astrophys. J. Lett.}}

\newcommand{\grl}{    {\it Geophys. Res. Lett.}}

\newcommand{\jgr}{    {\it J. Geophys. Res.}}

\newcommand{\pasj}{   {\it Pub. Astron. Soc. Japan}}

\newcommand{\solphys}{{\it Solar Phys.}}
\newcommand{\sovast}{ {\it Soviet  Astron.}}
\newcommand{\ssr}{    {\it Space Sci. Rev.}}
\chardef\us=`\_

\begin{document}
\begin{article}
\begin{opening}

\title{Simulation of Quiet-Sun Hard X-rays Related to Solar Wind Superhalo Electrons}

\author[addressref=aff1]{\inits{W}\fnm{Wen}~\lnm{Wang}}
\author[addressref=aff1,corref,email={wanglhwang@gmail.com}]{\inits{L.}\fnm{Linghua}~\lnm{Wang}}
\author[addressref={aff2,aff3}]{\inits{S}\fnm{S\"am}~\lnm{Krucker}}
\author[addressref=aff4]{\inits{I}\fnm{Iain}~\lnm{Hannah}}

\address[id=aff1]{School of Earth and Space Sciences, Peking University, Beijing, 100871, China}
\address[id=aff2]{Space Sciences Laboratory, University of California at Berkeley, Berkeley, CA 94720, USA}
\address[id=aff3]{Institute of 4D Technologies, University of Applied Sciences Northwestern Switzerland, 5210 Windisch, Switzerland}
\address[id=aff4]{SUPA School of Physics and Astronomy, University of Glasgow, Glasgow, G12 8QQ, UK}

\runningauthor{Wang \etal}
\runningtitle{Quiet-Sun HXRs Related to Superhalo Electrons}

\begin{abstract}
In this paper, we propose that the accelerated electrons in the quiet Sun could collide with the solar atmosphere to emit Hard X-rays (HXRs) via non-thermal bremsstrahlung, while some of these electrons would move upwards and escape into the interplanetary medium, to form a superhalo electron population measured in the solar wind. After considering the electron energy loss due to Coulomb collisions and the ambipolar electrostatic potential, we find that the sources of the superhalo could only occur high in the corona (at a heliocentric altitude $\gtrsim 1.9$ R$_\odot$ (the mean radius of the Sun)), to remain a power-law shape of electron spectrum as observed by {\it Solar Terrestrial Relations Observatory} (STEREO) at 1AU near solar minimum \citep{Wang2012}. The modeled quiet-Sun HXRs related to the superhalo electrons fit well to a power-law spectrum, $f \sim \varepsilon^{-\gamma}$ in the photon energy $\varepsilon$, with an index $\gamma$ $\approx$ 2.0\textendash2.3 (3.3\textendash3.7) at 10\textendash100 keV, for the warm/cold thick-target (thin-target) emissions produced by the downward-traveling (upward-traveling) accelerated electrons. These simulated quiet-Sun spectra are significantly harder than the observed spectra of most solar HXR flares. Assuming that the quiet-Sun sources cover 5\% of the solar surface, the modeled thin-target HXRs are more than six orders of magnitude weaker than the {\it Reuven Ramaty High Energy Solar Spectroscopic Imager} (RHESSI) upper limit for quiet-Sun HXRs \citep{Hannah2010}. Using the thick-target model for the downward-traveling electrons, the RHESSI upper limit restricts the number of downward-traveling electrons to at most $\approx$3 times the number of escaping electrons. This ratio is fundamentally different from what is observed during solar flares associated with escaping electrons where the fraction of downward-traveling electrons dominates by a factor of 100 to 1000 over the escaping population.
\end{abstract}
\end{opening}

\section{Introduction}

The solar wind superhalo electron population is observed at energies above 2 keV even during very quiet times near 1 AU \citep{Lin98, Wang2012, Wang2015}, in the absence of any significant solar and interplanetary activities, \textit{e.g.}, solar active regions, flares, coronal mass ejections, radio bursts, stream interaction regions, {\it etc}. Using high-sensitivity electron measurements from {\it the 3D Plasma and Energetic Particle} (3DP) instrument on the WIND spacecraft and {\it the SupraThermal Electron} (STE) instrument on the STEREO spacecraft, \citet{Wang2012, Wang2015} found that the differential fluxes of superhalo electrons, $J(E)$, generally fit well to a power-law energy spectrum extending up to at least 200 keV with the form $J(E) \sim E^{-\beta}$, with an average $\beta \approx$ 2.3\textendash2.4. They suggested that this superhalo population could originate from escaping nonthermal electrons accelerated in the processes related to the source of the solar wind or coronal heating, \textit{e.g.}, continual microflaring and nanoflaring \citep[\textit{e.g.},][]{Parker1988}, or it could be produced by wave-particle interaction \citep[\textit{e.g.},][]{Yoon2011,Yoon2012a, Yoon2012b, Yoon2012c, Kim2015} and/or long-distance propagation effects in the interplanetary medium (IPM).

If the superhalo electrons originate from nonthermal processes at the Sun, then these upward-traveling electrons and their downward-traveling counterparts should collide with the solar atmosphere to generate Hard X-rays (HXRs) via nonthermal bremsstrahlung from the quiet Sun, \textit{i.e.}, areas free of active regions and solar flares. Based on the  magnetohydrodynamics (MHD) and test particle simulations, \citet{Yang2015} found that the electrons can be accelerated by the electric field generated by magnetic reconnection in the solar wind source region, to form a power-law energy spectrum with a spectral index of $\beta \approx$ 1.5\textendash2.4 at 2\textendash200 keV. In their simulations, about half of the accelerated electrons move upwards along the newly-opened magnetic field lines that may escape into the interplanetary space to form the superhalo population in the solar wind, while the other half move downwards to collide with the dense plasma. Since the 1960s, observations of the quiet Sun have provided upper limits on quiet-Sun HXRs \citep{Mandelshtam1965, Pet66, Edwards1967, Fef97}. Recently, \citet{Hannah2010} reported the best-to-date upper limit for quiet-Sun HXRs at 3\textendash200 keV, using high-sensitivity RHESSI measurements between July 2005 and April 2009.

In this paper, assuming interchange reconnection as the coronal source of superhalo electrons, we estimate the expected HXR emission generated in the corona/chromosphere via the warm/cold thick-target bremsstrahlung models and thin-target bremsstrahlung model for the STEREO measurements of quiet-time superhalo electrons near solar minimum. We also utilize a power-law shaped electron spectrum to constrain the altitude of the superhalo source at the Sun, after considering the electron energy loss en route to 1 AU.

\section{Simulations}

\begin{figure}
\includegraphics[width=1.0\textwidth]{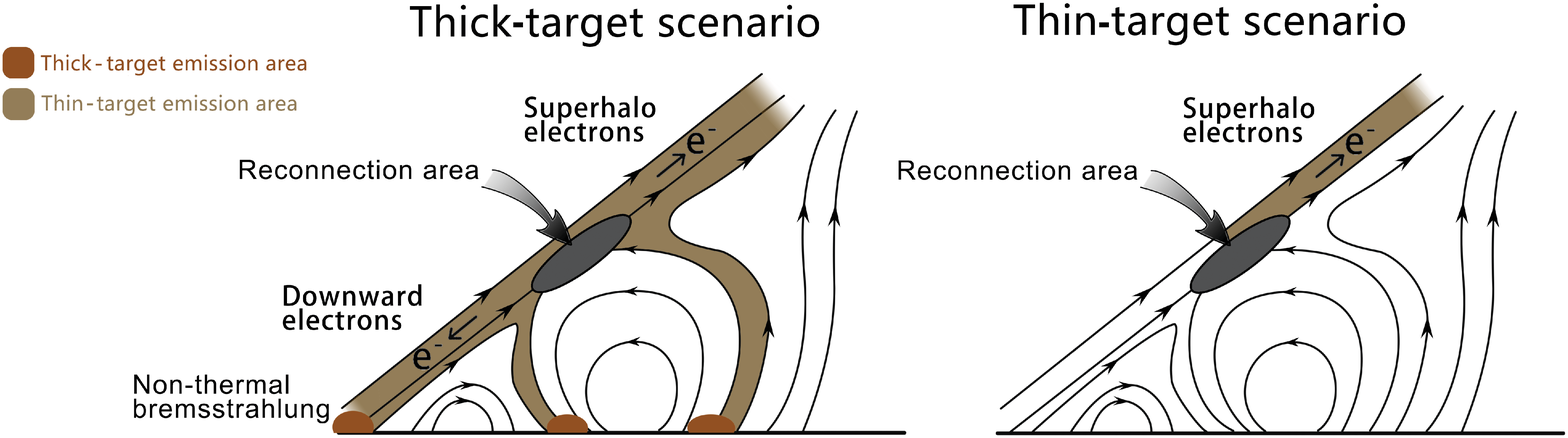}
\caption{2-D schematics of interchange reconnection, speculated for the source of solar wind superhalo electrons. These scenarios are similar to the flare model proposed by \citet{Shimojo2000}. Energized electrons escaping upwards from the acceleration site could form the superhalo population observed in the solar wind. The arrowed curves represent field lines. The thin-target scenario only involves the escaping energized electrons and their HXR emission, while the thick-target scenario can be appropriate for the HXR production in the presence of downward-traveling electrons.} \label{fig1}
\end{figure}

Inspired by \citet{Yang2015}, we propose a scenario of small interchange reconnections (that are probably related to the solar wind source or coronal heating, \textit{e.g.}, nanoflares) to associate the superhalo electrons in the interplanetary space with the HXR emissions in the quiet Sun (see Figure~\ref{fig1}). In this scenario, some of the accelerated electrons, \textit{e.g.}, by magnetic reconnection or turbulence, travel upwards along the open magnetic field lines into the interplanetary space, to form the solar wind superhalo population, while the rest propagate downwards into the lower atmosphere. Both these upward-traveling and downward-traveling electrons can collide with the solar atmosphere to generate the HXR emissions via nonthermal bremsstrahlung processes.

Regarding the energy loss of nonthermal particles to Coulomb collisions, there are two bremsstrahlung models: the thick-target model, where the accelerated particles lose all of their energies to emit HXRs, and the thin-target model, where the accelerated particles either escape from the interaction region before losing much energy or are continuously accelerated during interaction \citep{Ram86}. For the thick-target emissions, a cold-target approximation is dominated by the systematic collisional energy loss \citep[\textit{e.g.},][]{Brown1971}, while a warm-target model also includes the effects of collisional energy diffusion and thermalization of fast electrons \citep{Kon15}. Here we only consider the electron-proton bremsstrahlung process, since the electron-electron bremsstrahlung process is negligible at electron energies below $\approx$300 keV \citep{Koch1959, Hau75}.

After the electron acceleration in the source region, the upward-traveling electrons would emit weak HXRs via thin-target bremsstrahlung, while the downward-traveling electrons would produce weak thin-target HXRs through the tenuous corona and intense thick-target HXRs from the chromospheric footpoints/low corona (Figure~\ref{fig1}). During their propagation, the upward-traveling (escaping) electrons will lose energy to Coulomb collisions and to the ambipolar electrostatic potential \citep{Wan06}:
\begin{equation}
\begin{aligned}
  \frac{dE}{dr} &= (\frac{dE}{dr})_{\rm Coll} + (\frac{dE}{dr})_{\rm AEP} \\
        &= -1.82\times10^{-7}\frac{n(r)}{E} - \frac{0.994}{r^2},
\end{aligned}
\end{equation}
where $E$ is the electron energy in keV, $r$ is the heliocentric distance in R$_\odot$, and $n(r)$ is the plasma number density of the corona and solar wind in cm$^{-3}$ defined by \citet{Leb98} and \citet{Lem11}:
\begin{eqnarray}
n(r)&=&
\begin{cases}
2.99\times10^8r^{-16}+1.55\times10^8r^{-6}+3.6\times10^6r^{-1.5},  &\text{if $1.05 \leqslant r < 3$;}\\
3.74\times10^6r^{-2.14}+4.62\times10^8r^{-6.13},  &\text{if $3 \leqslant r < 5$;} \\
2.98\times10^6(r-0.9)^{-2.15}, &\text{if $5 \leqslant r$.}
\end{cases}
\end{eqnarray}

For a given electron energy channel $m$, we can obtain the upward-traveling electron flux $J_{\rm up}$ at the source altitude $r_0$ by assuming a constant electron flow \citep{Wan06}: $C_0J_{\rm up}\Delta E_{\rm 0m} = C_1J_{\rm 1m}\Delta E_{\rm 1m}$, where $J_{\rm up}$ (or $J_{\rm 1m}$) is the electron flux, $\Delta E_{\rm 0m}$ (or $\Delta E_{\rm 1m}$) is the energy bandwidth, and $C_0 = 4\pi r_0^2$ (or $C_1= 4\pi (AU)^2$) is the cross-sectional area at $r_0$ (or at 1AU). For the 2\textendash20 keV power-law superhalo electrons observed at 1 AU \citep{Wang2012}, the derived source electrons at the Sun would have a power-law spectrum when $r_0 \gtrsim 1.9$ R$_\odot$, or have a spectrum that bends up at low energies when $r_0 < 1.9$ R$_\odot$.

In this study, we use a source altitude of $r_0$ = 1.9 R$_\odot$ to estimate the energy spectrum of upward-traveling electrons accelerated at the Sun, $J_{\rm up}(E)$, from the quiet-time observations of superhalo electrons by STEREO at 1 AU \citep{Wang2012}. A 2 keV (or 20 keV) electron detected at 1 AU thus begins as a 2.8 keV (or 20.5 keV) electron at 1.9 R$_\odot$. The 2-20 keV superhalo electrons at 1 AU correspond to the accelerated 2.8\textendash20.5 keV electrons at the source with a power-law spectral index, $\beta$, that ranges from $\approx$1.6 to $\approx$3.6 with a peak at 2.5\textendash3.0 (see Figure~\ref{fig2}a). The flux of downward-traveling accelerated electrons, $J_{\rm down}$, is assumed to be related to $J_{\rm up}$ by $J_{\rm down} = J_{\rm up}/\eta$, where $\eta$ is the number ratio between the upward-traveling and downward-traveling electrons in the source region.

In the absence of downward-traveling accelerated electrons, the HXRs would be produced by the upward-traveling accelerated electrons via thin-target bremsstrahlung process through the high corona, refered to as \textquotedblleft thin-target scenario\textquotedblright (see Figure~\ref{fig1}). In the presence of downward-traveling accelerated electrons, the HXR production would be dominated by thick-target emissions of downward-traveling electrons from the lower solar atmosphere, referred to as \textquotedblleft thick-target scenario\textquotedblright. According to the electron energy losses described in Equation (1), a 2.8 keV downward-traveling electron can only reach the low corona at altitudes $>1.1$ R$_\odot$, while a high-energy electron could reach the cold chromosphere. For the thick-target scenario, thus, a warm-target (or cold-target) model can be appropriate for the simulation of HXR production including (or excluding) the low-energy electrons of a few keV. In this study, we simulate the energy spectrum of HXRs generated by the downward-traveling electrons via the warm/cold thick-target bremsstrahlung model and by the upward-traveling electrons via the thin-target model in the Solar SoftWare (SSW) package \citep[\textit{e.g.},][]{Freeland1998}.

\subsection{Cold Thick-target Bremsstrahlung Model}
In the cold thick-target bremsstrahlung model, for a power-law energy spectrum of source (HXR-producing) electrons at the Sun,
 \begin{equation}
 J(E) = A \times \varepsilon^{-\beta},
 \end{equation}
the deduced energy spectrum of HXRs at the Sun can be described by the formula \citep{Brown1971}:
\begin{equation}
 f(\varepsilon) = 2.17 \int_\varepsilon^{E_{max}} \frac{S \cdot \sigma(\varepsilon,E) \cdot v^2(E)}{{\rm ln}\Lambda} \left[ \int_E^{E_{max}} J(E_0) \, dE_0 \right] \, dE,
\end{equation}
where $E$ and $\varepsilon$ denote the electron and photon energies, respectively, in keV, $v(E)$ is the electron velocity in cm s$^{-1}$, $J(E)$ and $f(\varepsilon)$ are the differential fluxes of electrons and photons, respectively, in units of keV$^{-1}$ s$^{-1}$ cm$^{-2}$, $S$ is the area of the radiating source region in cm$^2$, ${\rm ln}\Lambda$ is the Coulomb logarithm,  $E_{\rm max}$ is the high-energy cutoff of source electrons, and $\sigma(\varepsilon,E)$ is the bremsstrahlung cross section in cm$^2$ keV$^{-1}$, differential in photon energy $\varepsilon$ \citep{Hau97}. Using the non-relativistic Bethe-Heitler approximation to $\sigma(\varepsilon,E)$, the deduced HXR spectrum can be simplified to a power-law function \citep{Brown1971, Ram86}:
\begin{equation}
 f(\varepsilon)= \frac {3.4 \times 10^{-7}S\times A} {\gamma^2(\gamma-1)^3B(\gamma-\frac{1}{2},\frac{3}{2})}\times \varepsilon^{-(\beta-1)} \varpropto \varepsilon^{-\gamma},
\end{equation}
Where $B(p,q) = \int_0^1 u^{p-1}(1-u)^{q-1}du$ is the Beta-function. Note that the spectral index of HXRs, $\gamma$, is equal to $\beta-1$ for the cold thick-target bremsstrahlung using a non-relativistic approximation of $\sigma(\varepsilon,E)$.

For the SSW cold-thick-target model, there are four input parameters: $S$, $FD$ (the flux density of source electrons), $E_{\rm min}$ (the low-energy cutoff of source electrons) and $E_{\rm max}$. We set $E_{\rm min}$ to be 10 keV since the cold approximation could be valid only for the $>$ 10 keV source electrons \citep{Kon15}, and set $E_{\rm max}$ to be 1 MeV according to the energy of X-ray observations \citep{Hurford1974,Lin1974}. The electron flux density is calculated by integrating the downward-traveling electron spectrum over energy, $FD=\int_{E_{\rm min}}^{E_{\rm max}}J_{\rm down}(E)\,dE$. We assume that the quiet-Sun sources cover 5\% of the solar surface near solar minimum, \textit{i.e.}, $S = 5\%\times 2\pi R_{\rm S}^2 = 1.5\times10^{21}$ cm$^2$, similar to the average total area of coronal holes (that contain open magnetic field lines) measured by the {\it Extreme ultraviolet Imaging Telescope} (EIT) on {\it Solar and Heliospheric Observatory} (SOHO) between 2006-2010 \citep{Low10}.

\begin{figure}
\includegraphics[width=1.0\textwidth]{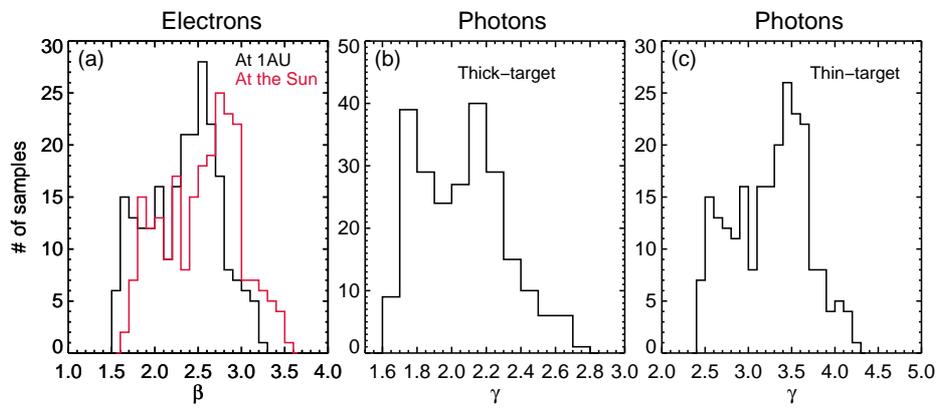}
\caption{Histograms of the power-law spectral index for the quiet-time superhalo electrons measured at 2\textendash20 keV by STEREO from March 2007 to March 2009 (a), the modeled quiet-Sun thick-target HXRs (b) and thin-target HXRs (c) at 10\textendash100 keV. The warm and cold thick-target models have the same results. } \label{fig2}
\end{figure}

For the STEREO quiet-time superhalo electron observations at 1AU \citep{Wang2012}, we calculate $J_{\rm up}$ at $r_0 = 1.9$ R$_\odot$ and simulate the quiet-Sun HXRs generated by $J_{\rm down}$ via the cold thick-target model. Figure~\ref{fig3}b plots the simulated cold-thick-target HXR spectra (black curves), related to the STEREO superhalo electrons with $\eta$ $=$ 0.33. To match the RHESSI quiet-Sun upper limits \citep{Hannah2010} requires a number ratio $\eta \gtrsim$ 33 \% between the upward-traveling and downward-traveling electrons in the solar source region. Figure~\ref{fig2}b plots the histogram of the power-law spectral index of cold-thick-target quiet-Sun HXRs, $\gamma$, fitted at energies of 10\textendash100 keV. Its major peak lies between 2.0 and 2.3, similar to the model prediction with the non-relativistic approximation ($\gamma = \beta-1$) from the estimated spectral indexes ($\beta \approx$ 2.5\textendash3.0) of the superhalo-related source electrons at 1.9 R$_\odot$ (see Figure~\ref{fig2}a). However, when $\beta$ $\lesssim$2, the bulk of HXRs are produced by the highest-energy electrons, so the modeled HXR spectrum is not sensitive to $\beta$ \citep{Kon06}, corresponding to the secondary $\gamma$ peak at 1.7\textendash1.8.

\subsection{Warm Thick-target Bremsstrahlung Model}
A warm-target bremsstrahlung model includes the effects of collisional energy diffusion and thermalization of fast electrons \citep{Kon15}. For the same HXR spectrum $f(\varepsilon)$, the deduced warm-target energy spectrum of source (HXR-producing) electrons at the Sun $J_{\rm wm}(E)$ (in units of keV$^{-1}$ s$^{-1}$ cm$^{-2}$) can be related to the modeled cold-target electron spectrum $J(E)$ by the formula \citep{Kon15}:
\begin{equation}
J_{\rm wm}(E) = 2.61\times 10^{-19} \cdot\frac{{\rm ln}\Lambda}{S} \frac{d}{dE} \left[ {G(\sqrt{\frac{E}{k_BT}})\cdot (\frac{J(E)}{E}(1-\frac{E}{k_BT})-\frac{dJ(E)}{dE}})\right],
\end{equation}
where $T$ is the average plasma temperature in K, $k_BT$ is in keV ($k_B$ is the Bolzmann's constant), $S$ is the area of radiating source region in cm$^2$ and $G(u)$ is the Chandrasekhar function \citep{Chandrasekhar1960}, given by
\begin{equation}
  G(u) = \frac{\rm erf({\it u})-{\it u}\cdot \rm erf'({\it u})}{2u^2},
\end{equation}
where ${\rm erf({\it u})} \equiv (2/\sqrt\pi)\int_0^u {\rm exp}(-t^2)\,dt$ is the error function.
For the SSW warm-thick-target model, there are 6 input parameters: $S$, $FD=\int_{E_{\rm min}}^{E_{\rm max}}J_{\rm down}(E)\,dE$, $E_{\rm min}$, $E_{\rm max}$, $\langle n \rangle$ (the average plasma density) and $T$. We set $E_{\rm min}$ = 2.8 keV, corresponding to an electron energy of 2 keV at 1 AU, and set $E_{\rm max}$ = 1 MeV according to the energy of X-ray observations \citep{Hurford1974,Lin1974}. We set $\langle n \rangle$ = $5\times 10^6$ cm$^{-3}$ according to the coronal density model used in this study (Equation (2)), and set $T$ to be $2\times 10^6$ K according to the nanoflare observations \citep{Schmelz2014,Aschwanden2000}.  The total area of radiating source region, $S$, is also estimated to be $5\%\times 2\pi R_{\rm S}^2 = 1.5\times 10^{21}$ cm$^2$.

For the 235 quiet-time samples of superhalo electrons observed by STEREO at 1AU near solar minimum \citep{Wang2012}, we calculate $J_{\rm up}$ at $r_0 = 1.9$ R$_\odot$ and simulate the quiet-Sun HXRs produced by $J_{\rm down}$ via the warm thick-target model. Figure~\ref{fig3}a plots the simulated HXR spectra (black curves) that are related to the STEREO superhalo electrons by $\eta$ = 0.33, compared with the upper limits of quiet-Sun HXRs observed by RHESSI near solar minimum (red arrows) \citep{Hannah2010}. The simulated HXR spectra appear to abruptly bend up at energies below $\approx$ 10 keV because of a pile-up of thermal electrons driven primarily by the effects of energy diffusion in the coronal target (E. Kontar, private communication), and to slightly bend down at energies above $\approx100$ keV. To match the RHESSI quiet-Sun upper limits, the number of upward-traveling (escaping) electrons aslo needs to be larger than $\approx$ 33 \% of the number of downward-traveling electrons in the quiet-Sun sources, \textit{i.e.}, $\eta \gtrsim 33\%$. Figure~\ref{fig2}(b) plots the histogram of the power-law spectral index of the warm-thick-target quiet-Sun HXRs, $\gamma$, fitted at energies of 10\textendash100 keV. It shows a major peak at 2.0\textendash2.3 and a secondary peak at 1.7\textendash1.8. This histogram is the same as that of the cold-thick-target HXR simulations since the HXR predictions from warm and cold thick-target models would differ only at energies below $\approx$ 10 keV (see Figure~\ref{fig3}).

\subsection{Thin-target Bremsstrahlung Model}
In the thin-target bremsstrahlung model, for a power-law spectrum of source electrons at the Sun as described by Equation (3), the deduced energy spectrum of HXRs at the Sun, $f(\varepsilon)$ (in units of keV$^{-1}$ s$^{-1}$ cm$^{-2}$), can be described by the formula \citep{Brown1971}:
\begin{equation}
 f(\varepsilon) = 2.37 \times 10^{35} \int_\varepsilon^{E_{max}} J(E) \times \sigma(\varepsilon,E) \, dE,
\end{equation}
where $J(E)$ is in units of keV$^{-1}$ s$^{-1}$ cm$^{-2}$, $\sigma(\varepsilon,E)$ is in cm$^{2}$ keV$^{-1}$, and $E$ is in keV. Using the non-relativistic Bethe-Heitler approximation to $\sigma(\varepsilon,E)$, the deduced HXR spectrum can also be simplified to a power-law function \citep{Ram86}:
\begin{equation}
 f(\varepsilon)= \frac{1.24 \times 10^{-24}A\langle n \rangle V} {\gamma(\gamma-1)^2B(\gamma-\frac{1}{2},\frac{3}{2})}\times {\varepsilon}^{-(\beta+1)} \varpropto \varepsilon^{-\gamma},
\end{equation}
where $V$ is the source region volume in cm$^{3}$. The spectral index of source electrons $\gamma$ is equal to $\beta+1$ for thin-target bremsstrahlung with the non-relativistic approximation to $\sigma(\varepsilon,E)$.

For the SSW thin-target model, there are 5 input parameters: $V$, $FD$, $E_{\rm min}$, $E_{\rm max}$ and $\langle n \rangle$. We set $E_{\rm min}$ (or $E_{\rm max}$) to be 2.8 keV (or 1 MeV), and set $FD$ to be the integral of the upward-traveling electron spectrum over energy. The average plasma density $\langle n \rangle$ is estimated to be $5\times 10^6$ cm$^{-3}$ at 1.9 R$_\odot$, according to the density model defined by Equation (2). We assume that the source depth, $h$, corresponds to an area $h^2$ matching the area of a single nanoflare, {\it i.e.}, $\approx 10^{15}$ cm$^{2}$ \citep{Tian2014}, so $h \approx 10^{7.5}cm$. For the thin-target quiet-Sun HXRs, the total volume of radiating source region is estimated to be $V = Sh$, with an area $S = 1.5\times10^{21}$ cm$^2$ (the same value used for thick-target emission).

Figure~\ref{fig3}c plots the simulated energy spectra of the thin-target quiet-Sun HXRs (black curves), related to the quiet-time STEREO superhalo observations. These thin-target HXR emissions are produced by the upward-traveling electrons in the absence of downward-traveling electrons, \textit{i.e.}, $\eta=\infty$. The modeled fluxes appear to be more than six orders of magnitude weaker than the RHESSI quiet-Sun upper limits near solar minimum \citep{Hannah2010}. The shown spectra can also fit well to a power-law function at energies from $\approx$5 keV up to 100 keV. At 10\textendash100 keV, the fitted spectral indices of thin-target quiet-Sun HXRs are mostly in the range 3.3\textendash3.7 (Figure~\ref{fig2}c), similar to the  non-relativistic thin-target prediction ($\gamma = \beta+1$) from the estimated spectral indexes of the superhalo-related source electrons at 1.9 R$_\odot$ (see Figure~\ref{fig2}a).

Figure~\ref{fig4} shows the integrated flux versus time for the thick-target (or thin-target) quiet-Sun HXRs over 10\textendash100 keV, modeled from the superhalo-related source electrons with a ratio of $\eta = 33\%$ ($\eta = \infty$). The warm and cold thick-target models have almost the same fluxes at energies above $\approx$ 10 keV. The correlation coefficient between the monthly average HXR flux (when available) and the monthly sunspot number is only 0.1\textendash0.3 for both thick-target and thin-target model results. Note that the thin-target quiet-Sun HXR fluxes are about seven orders of magnitude weaker than the thick-target HXR fluxes.

\section{Summary and Discussion}

In this study, we propose that in small interchange reconnections at the quiet Sun, the upward-traveling population of the accelerated electrons can escape into the interplanetary space to form the superhalo electrons measured in the solar wind, while the downward-traveling and/or upward-traveling populations can collide with the solar atmosphere to emit the quiet-Sun HXRs via nonthermal bremsstrahlung processes. After considering the electron energy loss due to Coulomb collisions and the ambipolar electrostatic potential in the corona and the IPM, we find that only the source electrons escaping from a heliocentric altitude of $r_0 \gtrsim 1.9$ R$_\odot$ can retain a power-law shape in the energy spectrum, to produce the 2\textendash20 keV power-law superhalo observed by STEREO at 1 AU near solar minimum \citep{Wang2012}. For the warm/cold thick-target (or thin-target) bremsstrahlung emissions, the modeled quiet-Sun HXRs have a power-law spectrum at energies of $\approx$10\textendash100 keV, with a spectral index $\gamma$ in the range 2.0\textendash2.3 (or 3.3\textendash3.7). On the other hand, the modeled thin-target and thick-target HXRs show no obvious solar cycle variation.

If the superhalo electrons come from escaping nonthermal electrons accelerated at the quiet Sun, their sources would lie high ($\gtrsim 1.9$ R$_\odot$) in the corona, in order to remain a power-law shape of energy spectrum as observed at 1 AU. The relevant thick-target quiet-Sun HXRs are modeled to have a power-law energy spectrum, $f(\varepsilon) \sim \varepsilon^{-\gamma}$, with an index $\gamma \approx$ 2.0\textendash2.3 at 10\textendash100 keV, significantly harder than the observed spectrum for most solar HXR flares \citep{Lin1974, Krucker2007}. This indicates that electron acceleration in the superhalo source region high in the corona (at low density) could be more efficient in accelerating high-energy electrons in relative terms than in the solar flare region in the low corona (at higher density). Our thick-target simulations also suggest that the number of upward-traveling (escaping) electrons is $\gtrsim 33\%$ of the number of downward-traveling electrons in the solar source region, to meet the upper limits for quiet-Sun HXRs observed by RHESSI \citep{Hannah2010}. Note that for thick-target models, the simulated number ratio of upward-travelling to downward-traveling electrons, $\eta$, is proportional to the quiet-Sun area, $S$. Here we assume that $S$ is similar to the total area of coronal holes that varies from 2.5\% to 7.5\% (5\% in average) of the solar surface, measured by SOHO/EIT between 2006-2010 \citep{Low10}. Thus, the number ratio $\eta$, varying from $\eta \gtrsim 17 \%$ to $\eta \gtrsim 50 \%$, is consistent with the test particle simulations of electron acceleration in the solar wind source region by \citet{Yang2015}, but it is more than one order of magnitude larger than the number ratio ($\approx 0.1\%-1\%$) estimated for solar HXR flares associated with solar energetic electron events \citep{Lin1974, Pan1984, Krucker2007}. Compared to solar flares that generally occur low in the corona, the superhalo sources likely lie high in the corona, leading to an easier escape of the accelerated electrons into the interplanetary space, \textit{i.e.}, a larger ratio of escaping electrons to downward-traveling electrons.

The thin-target quiet-Sun HXRs can be produced by the superhalo electrons through the high corona, and the modeled HXR fluxes are more than six orders of magnitude weaker than the RHESSI quiet-Sun upper limits. These thin-target HXRs also fit to a power-law energy spectrum with an index $\gamma \approx$ 3.3\textendash3.7 at 10\textendash100 keV, still harder than all but the solar flares with the hardest spectra \citep{Lin1974}. This also supports the possible presence of a relatively more efficient electron acceleration in the quiet Sun.

\begin{figure}
\includegraphics[width=1.0\textwidth]{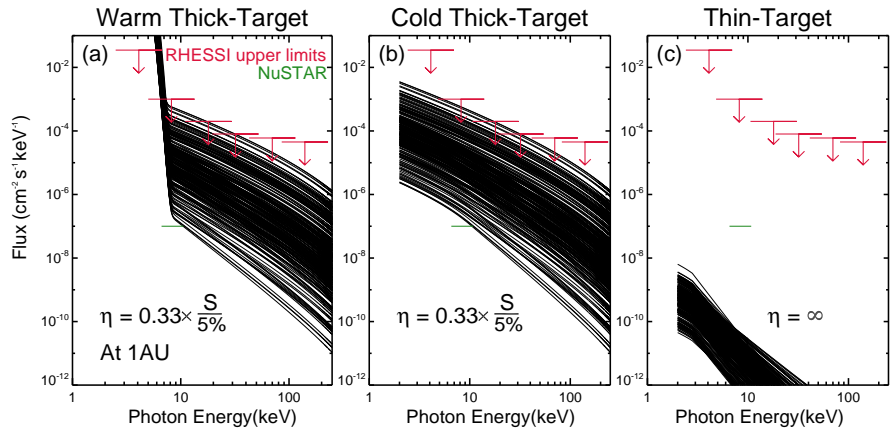}
\caption{The energy spectra of the modeled warm thick-target (a), cold thick-target (b) and thin-target (c) quiet-Sun HXRs, related to the 235 quiet-time samples of superhalo electrons measured by STEREO. The red arrows represent the upper limits of quiet-Sun HXRs measured by RHESSI \citep{Hannah2010}, and the green horizontal line shows the one-count limit at 7\textendash10 keV for an exposure of one hour for the {\it Nuclear Spectroscopic Telescope Array} (NuSTAR) mission. The number of downward-traveling electrons has been adjusted ($\eta \approx$ $0.33\times S/0.05$) so that their HXR emissions are just below the RHESSI upper limits. Note that the warm and cold thick-target models give almost the same HXR fluxes at energies above $\approx$ 10 keV.}\label{fig3}
\end{figure}

\begin{figure}
\includegraphics[width=1.0\textwidth]{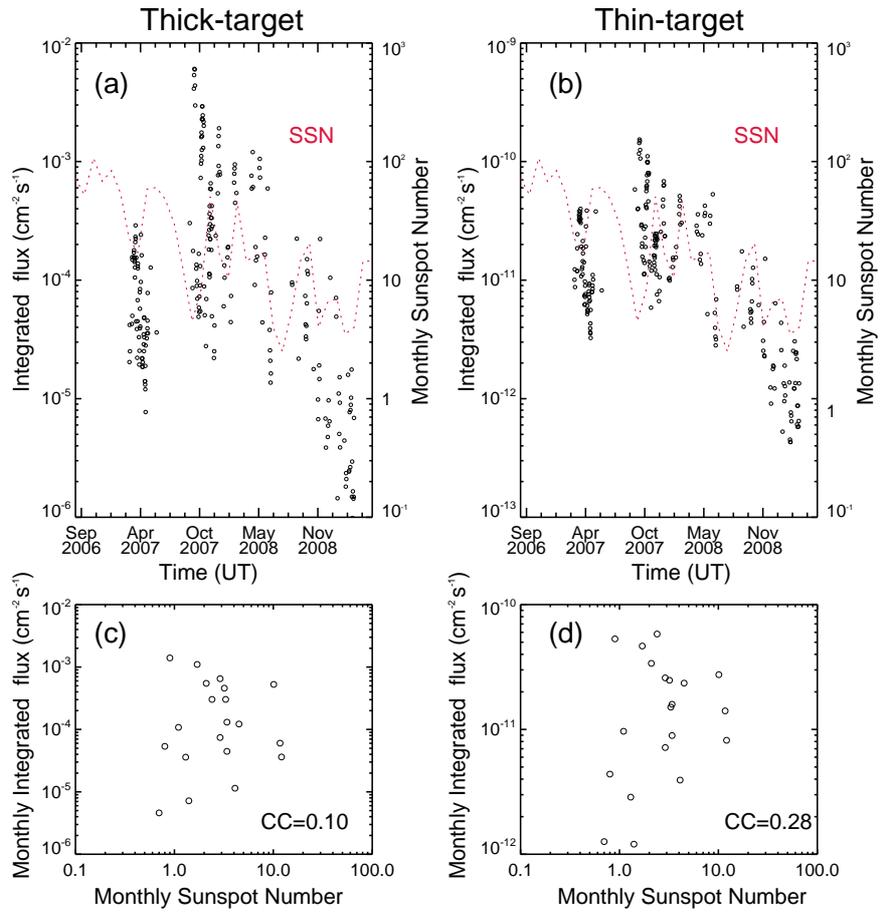}
\caption{Panels (a) and (b): The integrated fluxes of quiet-Sun HXRs over the energy range of 10\textendash100 keV, from the thick-target and thin-target models. The red dotted line represents the monthly sunspot number. Panels (c) and (d): Scatter diagrams of the monthly sunspot number versus the integrated flux of the monthly average modeled HXRs over 10\textendash100 keV.} \label{fig4}
\end{figure}

High-sensitivity measurements of quiet-Sun HXRs would greatly improve our understanding of the electron acceleration/transport at the Sun, especially for the generation of the solar wind superhalo electrons. However, previous HXR instruments were not sensitive enough so that they could only provide upper limits on the fluxes of quiet-Sun HXRs \citep[\textit{e.g.},][]{Fef97, Hannah2010}. Figure~\ref{fig3} shows the upper limits from RHESSI \citep{Hannah2010}, and the 1-count level at 7-10 keV for an exposure of 1 hour from NuSTAR. To estimate a detection threshold for NuSTAR is difficult as it strongly depends on the detector lifetime that is given by the other HXR sources within, but also outside the NuSTAR field-of-view \citep[\textit{e.g.},][]{Gref2016,Hannah2016}. In any case, NuSTAR should give a significant improvement relative to RHESSI. For the case that there is a downward component of similar number as the escaping population that forms the superhalo, NuSTAR has a chance of making a detection of the associated HXR emissions in the corona. However, it will be unlikely that a present-day or near-future instruments will observe the thin-target emission from escaping electrons. In this study, the simulated energy spectrum of quiet-Sun HXRs related to the quiet-time superhalo electrons can provide useful constraints, \textit{e.g.}, on the energy resolution and dynamic range, for the future instrumentation of the quiet-Sun observations.

The effects of wave-particle interaction in the IPM \citep[\textit{e.g.},][]{Yoon2011,Yoon2012a, Yoon2012b, Yoon2012c, Kim2015} are not taken into account in the present study. Instead of a solar source, \citet{Yoon2012c} and \citet{Kim2015} proposed that the superhalo electrons could be accelerated by local resonant interactions with (electrostatic) Langmuir waves excited by electrons during their progation through the IPM. Future investigations of the source of superhalo electrons and relevant HXRs will include a very careful and detailed modeling of interplanetary wave-particle interactions that exceeds the scope of this paper.

\begin{acks}
We thank Gordon Holman, Rui Liu and Eduard Kontar for helpful discussions. This research at Peking University is supported in part by NSFC under contracts 41421003, 41274172, 41474148, 41231069. I.G. Hannah acknowledges support from a Royal Society University Research Fellowship. S. Krucker is supported by Swiss National Science Foundation (200021-140308) and through NASA contract NAS 5-98033 for RHESSI.
\end{acks}


\clearpage

\clearpage

\clearpage

\clearpage

\end{article}
\end{document}